\documentclass[aps,twocolumn,showpacs,psfig,superscriptaddress,longbibliography]{revtex4-2}
\usepackage{amsfonts}
\usepackage{mathrsfs}
\usepackage{amsmath}
\usepackage{color}
\usepackage{natbib}
\usepackage{textcomp}
\usepackage{graphicx}
\usepackage{bm}
\usepackage{amssymb}

\usepackage{xspace}
\usepackage{epstopdf}
\usepackage{dcolumn}
\usepackage{longtable}
\usepackage{multirow}
\usepackage[colorlinks=true, letterpaper=true, pdfstartview=FitV, linkcolor=blue, citecolor=blue, urlcolor=blue]{hyperref}
\usepackage{float}

\makeatletter

\newcommand{\Rmnum}[1]{\expandafter\@slowromancap\romannumeral #1@}
\makeatother

\begin{document}

\title{Diverse Manifestations of Electron-Phonon Coupling in a Kagome Superconductor}

 \author{Jing-Yang You}
 \affiliation{Mork Family Department of Chemical Engineering and Materials Science, University of Southern 
California, Los Angeles, CA 90089, USA}
 \affiliation{Department of Physics, National University of Singapore, 2 Science Drive 3, Singapore 117551, Singapore}

 \author{Chih-En Hsu}
 \affiliation{Mork Family Department of Chemical Engineering and Materials Science, University of Southern 
California, Los Angeles, CA 90089, USA}
 \affiliation{Department of Physics, Tamkang University, Tamsui, New Taipei 251301, Taiwan}

 \author{Mauro Del Ben}
\affiliation{Applied Mathematics and Computational Research Division, Lawrence Berkeley National Laboratory, Berkeley, California 94720, USA}

\author{Zhenglu Li}
\email{zhenglul@usc.edu}
\affiliation{Mork Family Department of Chemical Engineering and Materials Science, University of Southern 
California, Los Angeles, CA 90089, USA}

\begin{abstract}
Recent angle-resolved photoemission spectroscopy (ARPES) experiments on a kagome metal CsV$_3$Sb$_5$  revealed distinct multimodal dispersion kinks and nodeless superconducting gaps across multiple electron bands. The prominent photoemission kinks suggest a definitive coupling between electrons and certain collective modes, yet the precise nature of this interaction and its connection to superconductivity remain to be established.
Here, employing the state-of-the-art \textit{ab initio} many-body perturbation theory computation, we present direct evidence that electron-phonon ($e$-ph) coupling induces the multimodal photoemission kinks in CsV$_3$Sb$_5$, and profoundly, drives the nodeless $s$-wave superconductivity, showcasing the diverse manifestations of the $e$-ph coupling.
Our calculations well capture the experimentally measured kinks and their fine structures, and reveal that vibrations from different atomic species dictate the multimodal behavior. Results from anisotropic $GW$-Eliashberg equations predict a phonon-mediated superconductivity with nodeless $s$-wave gaps, in excellent agreement with various ARPES and scanning tunneling spectroscopy measurements. 
Despite of the universal origin from the $e$-ph coupling, the contributions of several characteristic phonon vibrations vary in different phenomena, highlighting a versatile role of $e$-ph coupling in shaping the low-energy excitations of kagome metals. 
\end{abstract}

\maketitle

\begin{figure}[!htbp]
  \centering
  \includegraphics[scale=0.5,angle=0]{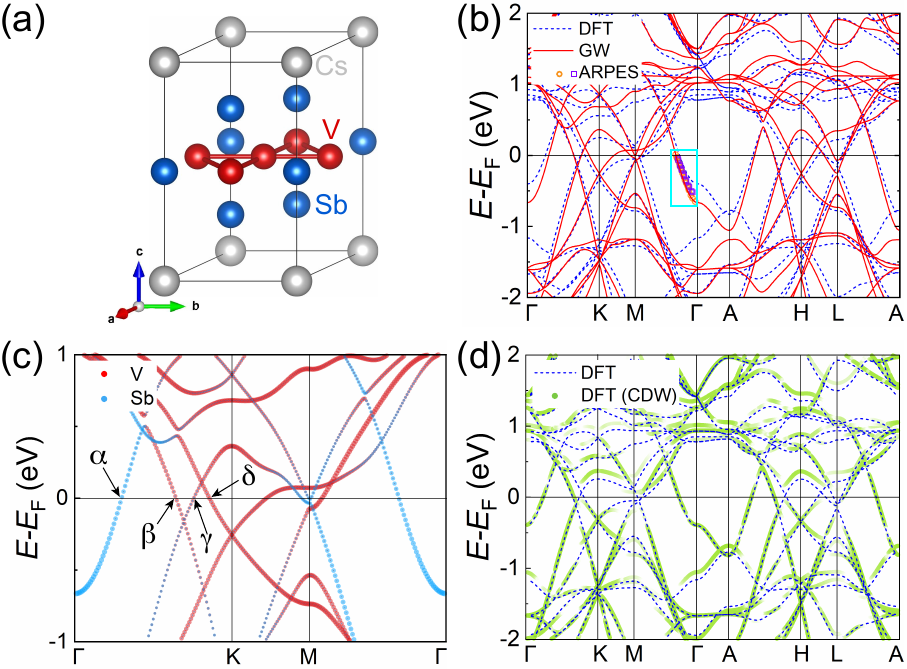}\\
  \caption{(a) Crystal structure of CsV$_3$Sb$_5$. (b) Comparison of DFT bands (blue dashed lines), $GW$ bands (red solid lines), and ARPES measurements (orange circles~\cite{Kang2022} and purple squares~\cite{Luo2023}). (c) Atomic orbital-projected $GW$ band structure, with four bands labeled as $\alpha$-, $\beta$-, $\gamma$- and $\delta$-bands. (d) Comparison of the band structure of the pristine phase and unfolded band structure (spectral function) for the CDW phase at the DFT level.} \label{fig1}
\end{figure}

Vanadium-based kagome metals $A$V$_3$Sb$_5$ ($A$ = K, Rb, Cs) exhibit a variety of intertwined phases, including superconductivity~\cite{Ortiz2020,Chen2021}, charge density wave (CDW)~\cite{Zhao2021,Liang2021,Li2021,Jiang2022}, nematicitic phases~\cite{Jiang2021,Xiang2021,Li2022,asaba2024evidence}, and topological orders~\cite{Ortiz2020,Hu2022}. In CsV$_3$Sb$_5$ [Fig.~\ref{fig1}(a)], intriguing superconductivity arises with a critical temperature ($T_c$) $\sim 2.5$ K under ambient pressure~\cite{Chen2021,Liang2021,Xu2021,Duan2021,Gupta2022,Gupta2022a,He2022,Roppongi2023,Zhang2023}, coexisting with the CDW order. In the presence of pressure and isovalent substitutional doping~\cite{Yang2022,Li2022a,Zhong2023a,Deng2024} [e.g., Cs(V$_{1-x}$Ta$_x$)$_3$Sb$_5$], the CDW order can be suppressed and the superconducting $T_c$ is enhanced to $\sim 4 - 5$ K.

Despite a number of different proposed superconducting pairing symmetries~\cite{Chen2021,Liang2021,Xu2021,Roppongi2023,Duan2021,Zhang2023,Zhao2021a,Ni2021,Mu2021,Zhao2021b,Wu2021}, recent angle-resolved photoemission spectroscopy (ARPES)~\cite{Zhong2023a} experiments revealed clear signatures of nodeless superconducting gaps in doped CsV$_3$Sb$_5$ systems, imposing strong constraints on the paring mechanisms. Moreover, multimodal photoemission kinks were observed \cite{Zhong2023,wu2023unidirectional,Luo2023}, exhibiting distinct behaviors between the different electron bands, with one showing a dominantly single kink and another displaying double kinks. Understanding the relation and origin of these low-energy excitations is crucial for disentangling the intertwined orders of kagome metals. However, a direct connection between photoemission kinks and other electronic orders, particularly superconductivity, is not always straightforward. A salient example is the strongly correlated unconventional copper-oxide superconductor, where the electron-phonon ($e$-ph) coupling responsible for the $\sim$70 meV photoemission kinks is deemed too weak to account for the high $T_c$ and not compatible with the $d$-wave pairing symmetry~\cite{lanzara2001evidence,sobota2021angle,Li2021a}.

A predictive first-principles investigation with appropriate level of theories can provide detailed descriptions of the interaction landscape. In CsV$_3$Sb$_5$, the low-energy photoemission kinks appear at $\sim$10$-$40 meV binding energy, which is compatible with the phonon vibrational frequencies~\cite{Liu2022a}. It is thus natural to  investigate the role of $e$-ph coupling in the low-energy phenomena of CsV$_3$Sb$_5$, as suggested by Ref.~\cite{Zhong2023}. Early first-principles calculations based on density functional theory (DFT) and density-functional perturbation theory (DFPT) suggested a weak to moderate overall $e$-ph coupling for CsV$_3$Sb$_5$ under ambient and elevated pressures~\cite{Zhang2021,Wang2023,Tan2021}. However, the precise contribution of $e$-ph coupling to various low-energy excitations remains elusive, partly due to a lack of direct theoretical studies of the experimentally relevant spectroscopic properties. Additionally, growing evidence \cite{yin2013correlation,antonius2014many,Li2019,Li2021a,Li2024a} has pointed out an inadequate description of the exchange-correlation effects in standard static DFT  approaches for $e$-ph coupling in certain materials. The recent development of  $GW$ perturbation theory ($GW$PT)~\cite{Li2019,Li2021a,Li2024,Li2024a}  enabled accurate descriptions of $e$-ph properties while
capturing important self-energy effects beyond DFT and DFPT, providing an excellent opportunity to elucidate the intriguing $e$-ph coupling in CsV$_3$Sb$_5$ from first principles.

In this work, using state-of-the-art \textit{ab initio} $GW$-based many-body perturbation theory approaches, we reveal that the $e$-ph coupling is the universal origin of the multimodal photoemission kinks and nodeless $s$-wave superconductivity in CsV$_3$Sb$_5$. Our calculated spectral functions demonstrate excellent agreement with experimental measurements, highlighting distinctive behaviors in the phonon-induced photoemission kinks in different electron bands of Sb and V characters, respectively. The origin of multimodal kinks can be readily attributed to phonon vibrations from different atoms with disparate atomic masses. The resemblance with ARPES experiments definitively calibrates the $GW$PT-calculated $e$-ph matrix elements, which are the scattering amplitudes between electron states (band $n$ and wavevector $\textbf{k}$) connected by different phonon modes (branch $\nu$ and wavevector $\textbf{q}$), as well as the $GW$ band structure. With the same set of microscopic $e$-ph coupling ingredients of CsV$_3$Sb$_5$, our fully anisotropic (\textbf{k}- and $n$-dependent) $GW$-Eliashberg calculations predict a phonon-mediated superconductivity with $T_c = 4.1$ K, consistent with the experimental values~\cite{Chen2021,Liang2021,Xu2021,Duan2021,Gupta2022,Gupta2022a,He2022,Roppongi2023,Zhang2023}. Moreover, the calculated superconducting gaps on multiple Fermi surfaces show an $s$-wave nature, agreeing well with ARPES-extracted nodeless gap distributions \cite{Zhong2023a} and measured superconducting quasiparticle density of states (DOS) from scanning tunneling spectroscopy (STS) experiments~\cite{Xu2021,Liang2021,Chen2021,Deng2024}.
Our first-principles findings unambiguously establish a direct connection between the multimodal kinks and the nodeless superconductivity of CsV$_3$Sb$_5$, elucidating their universal origin with diverse vibrational characteristics stemming from the $e$-ph coupling.

Figure~\ref{fig1}(b) presents the electron band structures of CsV$_3$Sb$_5$ calculated using both DFT and $GW$ approaches. Multiple bands intersect the Fermi level ($E_\text{F}$), which is set to zero. The $GW$ self-energy effects mainly shift the electron pocket around $\Gamma$ downward, positioning the corresponding band energy at $\Gamma$ to be $-0.63$ eV ($-0.40$ eV) at the $GW$ (DFT) level, in good agreement with the experimental value of $-0.6 \sim -0.68$ eV~\cite{Kang2022,Luo2023}. The overall difference between DFT and $GW$ band structures is minor (the self-energy renormalization in $e$-ph coupling is not strong either, see Supplemental Material Fig. S1), indicating relatively weak correlation effects (captured by $GW$ self-energy) in CsV$_3$Sb$_5$, consistent with prior knowledge \cite{Ortiz2020,Zhao2021b,Jiang2022,Liu2022}. 
Figure~\ref{fig1}(c) shows the $GW$ band structure near $E_\text{F}$ with orbital projections onto Sb states (blue) and V states (red), with Cs states having negligible presence within this energy range. Four bands near  $E_\text{F}$ are labeled as $\alpha$, $\beta$, $\gamma$, and $\delta$ for further analysis and discussions. 
Spin-orbit coupling (SOC) can possibly affect the properties of materials~\cite{Heid2010}. In the case of CsV$_3$Sb$_5$, our explicit calculations show that SOC has negligible effects on the electronic states near the $E_\text{F}$ and on the $e$-ph coupling strength (see Fig. S2 and more discussions in Supplemental Material). Hence, we will continue our subsequent discussions without further considering SOC.

The CDW phase of vanadium-based kagome metals is a central topic of research, because of the experimental implications, as well as intensive debates, on the possible chiral charge order~\cite{Jiang2021,Ortiz2021,Guo2022,Xiao2023,Feng2023} and time-reversal symmetry breaking~\cite{Xu2022,Zhang2023,Farhang2023,Saykin2023,Deng2024,Deng2024a}.
In this work, we focus on the manifestations of $e$-ph coupling in CsV$_3$Sb$_5$, thus it is intriguing to disentangle the interplay between CDW and superconductivity, which coexist in undoped compounds under ambient pressure~\cite{Chen2021,Zhao2021,Liang2021}.
Fig. 1(d) shows the electronic structure (the spectral function intensity map) of CsV$_3$Sb$_5$ in the CDW phase, unfolded back to the Brillouin zone of the primitive unit cell, comparing with the band structure of the pristine phase (without CDW), at the DFT level.
Despite the structural distortions, the overall band structure of the CDW phase remains largely similar to that of the pristine phase, except for some band-gap openings near $E_\text{F}$, thus reducing the DOS at $E_\text{F}$ and suggesting a suppression effect on the $e$-ph coupling and superconductivity (see Supplemental Material for more discussions on the CDW phase).

In the following discussions, we will firstly and mostly focus on the intrinsic $e$-ph coupling and its consequences in CsV$_3$Sb$_5$ without being obscured by the CDW structural distortions. These results are directly relevant to the pristine phase of CsV$_3$Sb$_5$ under pressure~\cite{Yu2021,Yu2022} or with isovalent substitutional doping~\cite{Yang2022,Li2022a,Zhong2023a,Deng2024}, as well as to the electron bands that are nearly unaffected by the CDW phase (e.g., the $\alpha$- and $\beta$-bands). 
We will rewind back to the discussions about the impact of the CDW phase, after a comprehensive understanding of the intrinsic $e$-ph coupling and superconducting properties of CsV$_3$Sb$_5$ is established. (For well-defined $e$-ph coupling calculations in the primitive cell, stable phonons are needed and can be created, for CsV$_3$Sb$_5$, by using norm-conserving pseudopotentials, which are also required by $GW$ and $GW$PT calculations. See Supplemental Material for more details.)

\begin{figure*}[!htbp]
  \centering
  \includegraphics[scale=0.4,angle=0]{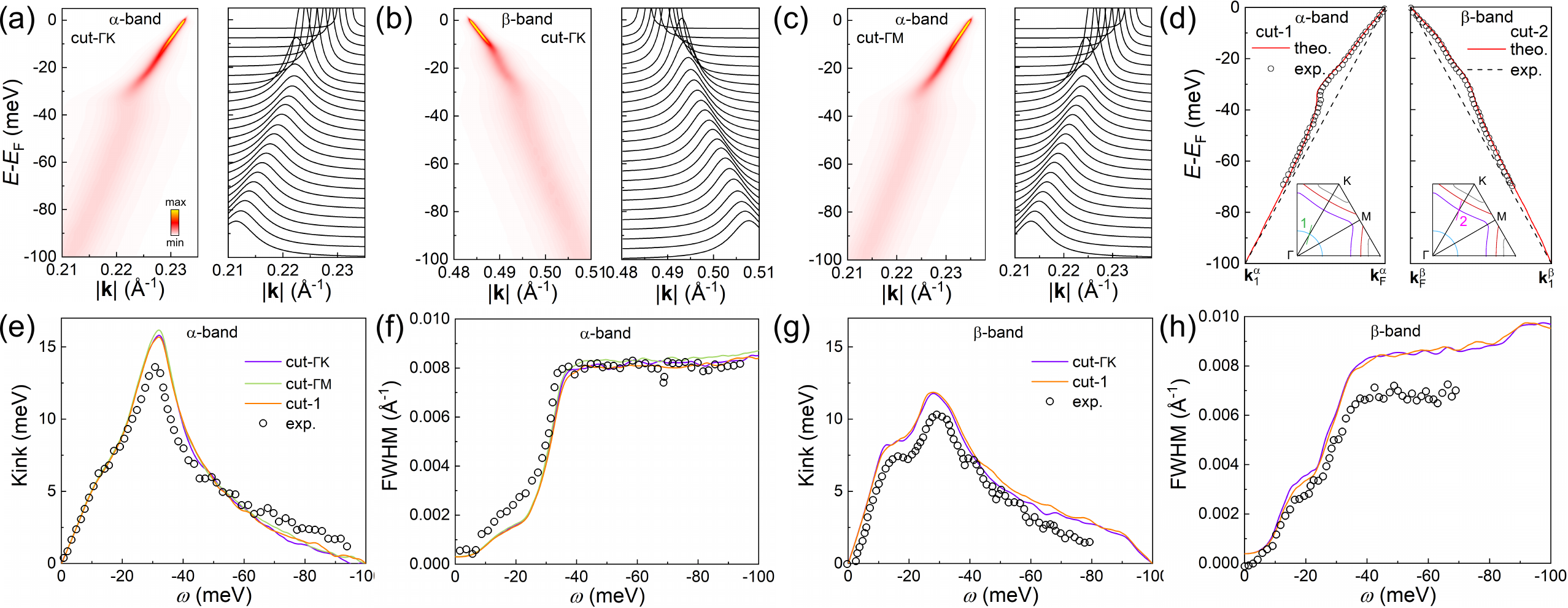}\\
  \caption{Spectral functions $A_{n\mathbf{k}}(\omega)$ (plotted in color scale) calculated with $e$-ph coupling included at $T$ = 6 K, and the corresponding momentum distribution curves (MDCs) between $E_{\rm F}$ and $E_1 = -100$ meV, for (a) $\alpha$- and (b) $\beta$-bands along the $\Gamma-K$ path, and (c) for $\alpha$-band along the $\Gamma-M$ path, respectively. Each MDC is fitted to a Lorentzian function to obtain the peak position and linewidth. (d) Comparison of the MDC-derived dispersion relation between experiment (open circles)~\cite{Zhong2023} and theory (red line) along cut-1 and cut-2 (as defined in insets), with the reference lines (black dashed lines) aligned. $\mathbf{k}_{\rm F}$ is the Fermi wavevector and $\mathbf{ k}_1$ is the wavevector corresponding to $E_1$. (e) Kink size as a function of quasiparticle state energy along different cuts, extracted as the energy difference between the MDC-derived dispersion relation and the straight reference line, for $\alpha$-band. (f) FWHM linewidth of the MDCs due to $e$-ph coupling. In (e) and (f), the experimental data \cite{Zhong2023} are from a cut parallel to and near cut-$\Gamma K$ (see Supplemental Material). (g) and (h) Similar to (e) and (f) but for $\beta$-band.}\label{fig2}
\end{figure*}

We now explore in detail the electron dispersion relation in the low-energy ($<$ 40 meV binding energy) region, where ARPES experiments show clear photoemission kinks, indicating evident electron-boson coupling. 
Our focus lies particularly on the $\alpha$-band (mainly of Sb characters) and $\beta$-band (mainly of V characters), as labeled in Fig.~\ref{fig1}(c). Figures~\ref{fig2}(a) and (b) display the calculated electron spectral functions $A_{n\mathbf{k}}(\omega)$ with $e$-ph interaction included \cite{Li2021a,giustino2008small}, and the corresponding momentum distribution curves (MDCs), for the $\alpha$- and $\beta$-bands along the $\Gamma$$-$$K$ Brillioun-zone cut, respectively. Clear signatures of phonon-induced electron self-energy effects including the dispersion kink and spectral width broadening show up. Notably, the $\alpha$-band displays a single kink around $-32$ meV, whereas the $\beta$-band shows double kinks at $-12$ and $-30$ meV, respectively, agreeing well with the experimental measurements~\cite{Zhong2023}. 
Another cut along the $\Gamma$$-$$M$ path reveals the similar spectral function in the $\alpha$-band as shown in Fig.~\ref{fig2}(c). For direct comparison with the experimental data, we also calculate spectral functions along two specific cuts (cut-1 and cut-2) adopted in Ref.~\cite{Zhong2023} (see Fig. S4). Following the same data processing procedure as in experiments~\cite{Zhong2023}, we fit MDCs to Lorentzian functions to extract the MDC-derived energy $vs$ wavevector dispersion relations and linewidth information [full width at half maximum (FWHM)]. Typically, the bare band (i.e., the non-interacting band) information in experiments is unknown.
Therefore, a common practice involves introducing a straight reference line by connecting two points on the dispersion relation (points at $E_\text{F}$ and $-0.1$ eV in Ref.~\cite{Zhong2023} and this work) to facilitate the extraction of many-body effects and fine features within a certain energy range. By aligning the reference lines for our and the ARPES data~\cite{Zhong2023}, Fig.~\ref{fig2}(d) directly compares the theoretical and experimental electron dispersion relations along cut-1 and cut-2, showing excellent agreement. 

Figure~\ref{fig2}(e) shows the kink magnitude (energy difference between the dispersion relation and reference line) as a function of the $\alpha$-band energy. A prominent peak at $\sim$32 meV is observed, with a subtle shoulder structure around 12 meV, indicating the presence of two modes coupling to electrons. The multimodal coupling in the $\alpha$-band is more obvious in the imaginary part of the $e$-ph self-energy, as manifested by two distinct slopes (in the energy range $\omega < 40$ meV) in the FWHM from the MDC analysis, as shown in Fig.~\ref{fig2}(f). Nevertheless, the $-12$ meV mode remains largely obscured by the main kink at $-32$ meV in the $\alpha$-band. 
Examination of the dispersion kinks along different Brillouin-zone cut directions (in particular, $\Gamma$$-$$K$, $\Gamma$$-$$M$, and cut-1) reveals directional isotropy, consistent with the ARPES conclusion~\cite{Zhong2023}. Figures~\ref{fig2}(g) and \ref{fig2}(h) illustrate the kinks and linewidths of the $\beta$-band. While the overall kink size appears weaker, a clear double-kink structure emerges, highlighting the multi-phonon mode coupling nature in CsV$_3$Sb$_5$. 

\begin{figure}[!htbp]
  \centering
  \includegraphics[scale=0.42,angle=0]{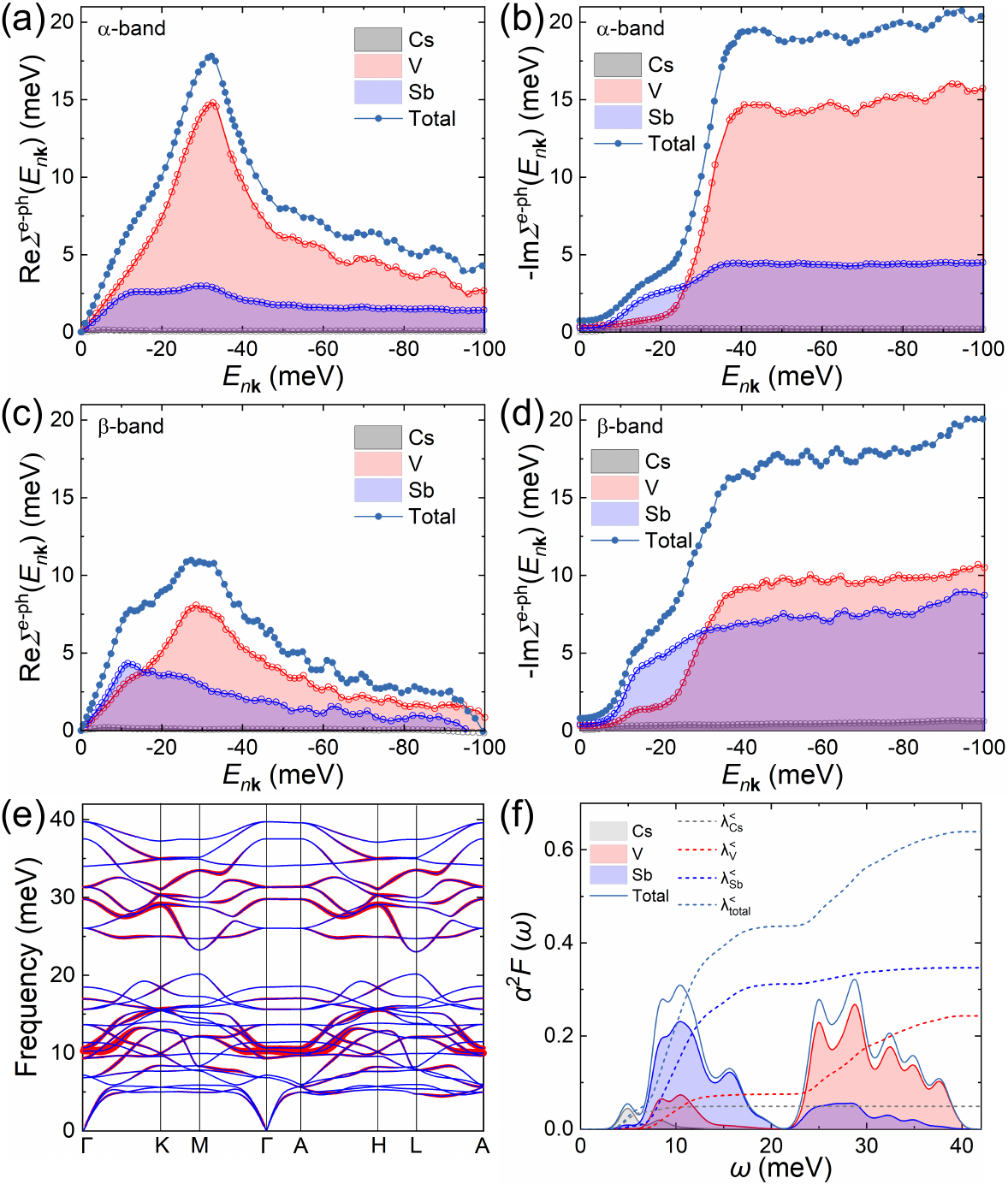}\\
  \caption{Contribution of $e$-ph coupling to the real part of the electron self-energy Re$\Sigma_{n\mathbf{k}}^{e-{\rm ph}}(\omega)$ from different atom vibrations for (a) $\alpha$- and (c) $\beta$-bands at $T$ = 6 K. Imaginary part of the electron self-energy Im$\Sigma_{n\mathbf{k}}^{e-{\rm ph}}(\omega)$ decomposed to atom vibrations for (b) $\alpha$- and (d) $\beta$-bands at $T$ = 6 K. (a)-(d) present data along the $\Gamma$$-$$K$ direction. (e) Phonon spectrum overlaid by the phonon-mode resolved $e$-ph coupling strength $\lambda_{\mathbf{ q}\nu}$. (f) Total and atom-vibration-decomposed Eliashberg spectral function $\alpha^2F(\omega)$ and cumulative $\lambda^{<}(\omega)$.}\label{fig3}
\end{figure}

To elucidate the origin of the multimodal kinks, we further calculate the atom-vibration resolved real and imaginary parts of the $e$-ph self-energy for the $\alpha$- and $\beta$-bands, respectively, as shown in Figs.~\ref{fig3}(a)-(d). For the $\alpha$-band, the kink at $-32$ meV is primarily attributed to V atom vibrations. Although there is a minor peak at $-12$ meV contributed by Sb atom vibrations, its intensity is overshadowed by the stronger signal from V vibrations, resulting in a single dominant kink [Fig.~\ref{fig3}(a)], as is evident in both our calculations and experiments~\cite{Zhong2023}. However, for the $\beta$-band, the weaker $e$-ph coupling strength and self-energy magnitude highlight the low-energy mode, resulting in two kinks [Fig.~\ref{fig3}(c)]: one mainly contributed by Sb vibrations at $-$12 meV, and the other mainly contributed by V vibrations at $-$30 meV. 
The atomic vibrational contributions to the spectral functions (Fig. S5) exhibit similar findings. As commonly practiced in ARPES experiments, the $e$-ph coupling strength can be estimated from the photoemission kinks (we denote here as $\lambda^\Sigma$)~\cite{Hofmann2009} with $\lambda^\Sigma=-\partial {\rm Re}\Sigma(\omega)/\partial \omega \vert_{\omega=E_{\rm F}}$, at sufficiently low temperatures. The estimated $\lambda^\Sigma$ (the kink-estimated $e$-ph coupling strength) values for the $\alpha$- and $\beta$-bands are 0.56 and 0.61, respectively, based on our calculated spectral functions.

It is critical to explore the implications of $e$-ph coupling for the superconductivity of CsV$_3$Sb$_5$, given that our calculated underlying $e$-ph coupling excellently agrees with and explains the experimental photoemission kinks~\cite{Zhong2023}. Fig.~\ref{fig3}(e) reveals two distinct groups of phonon bands separated by a gap at around 20$-$23 meV, both displaying strong coupling as dictated by the phonon-mode decomposed $e$-ph coupling $\lambda_{\textbf{q}\nu}$. 
The projected phonon DOS (Fig.~S6) shows that Cs atom vibrations dominate the lowest frequency range ($\sim$3$-$8 meV), Sb atom vibrations are primarily concentrated in the intermediate frequency range (mainly $\sim$4$-$20 meV, with minor projection hybridization in the high-energy range), and V atom vibrations are essentially situated in the high-frequency region (mainly $\sim$23$-$40 meV, with minor projection hybridization in the intermediate-energy range). This clear frequency separation can be attributed to their distinct atomic masses, with $m_\text{Cs} > m_\text{Sb} > m_\text{V}$. Figure~\ref{fig3}(f) plots the Eliashberg function $\alpha^2F(\omega)$ and the cumulative $e$-ph coupling strength $\lambda^< (\omega) = 2\int_0^\omega \frac{\alpha^2F(\omega')}{\omega'}d\omega'$. 
A total $\lambda$ of 0.64 is obtained, with Sb vibrations contributing 54\%, V vibrations contributing 38\%, and Cs vibrations contributing 8\% to the total value, respectively. The relative vibrational contributions to the overall $\lambda$ contrast with the photoemission kinks, where V vibrations exhibit stronger coupling. This distinct behavior arises because the relevant electron states are at different excitation energies. In photoemission kinks, electrons at $E_\text{F}-\omega_\text{ph}$ couple the most to the corresponding phonon modes with the frequency $\omega_\text{ph}$, whereas in superconductivity, the coupling with phonon modes of lower frequencies weighs more for electrons at $E_\text{F}$.

\begin{figure}[!htbp]
  \centering
  \includegraphics[scale=0.38,angle=0]{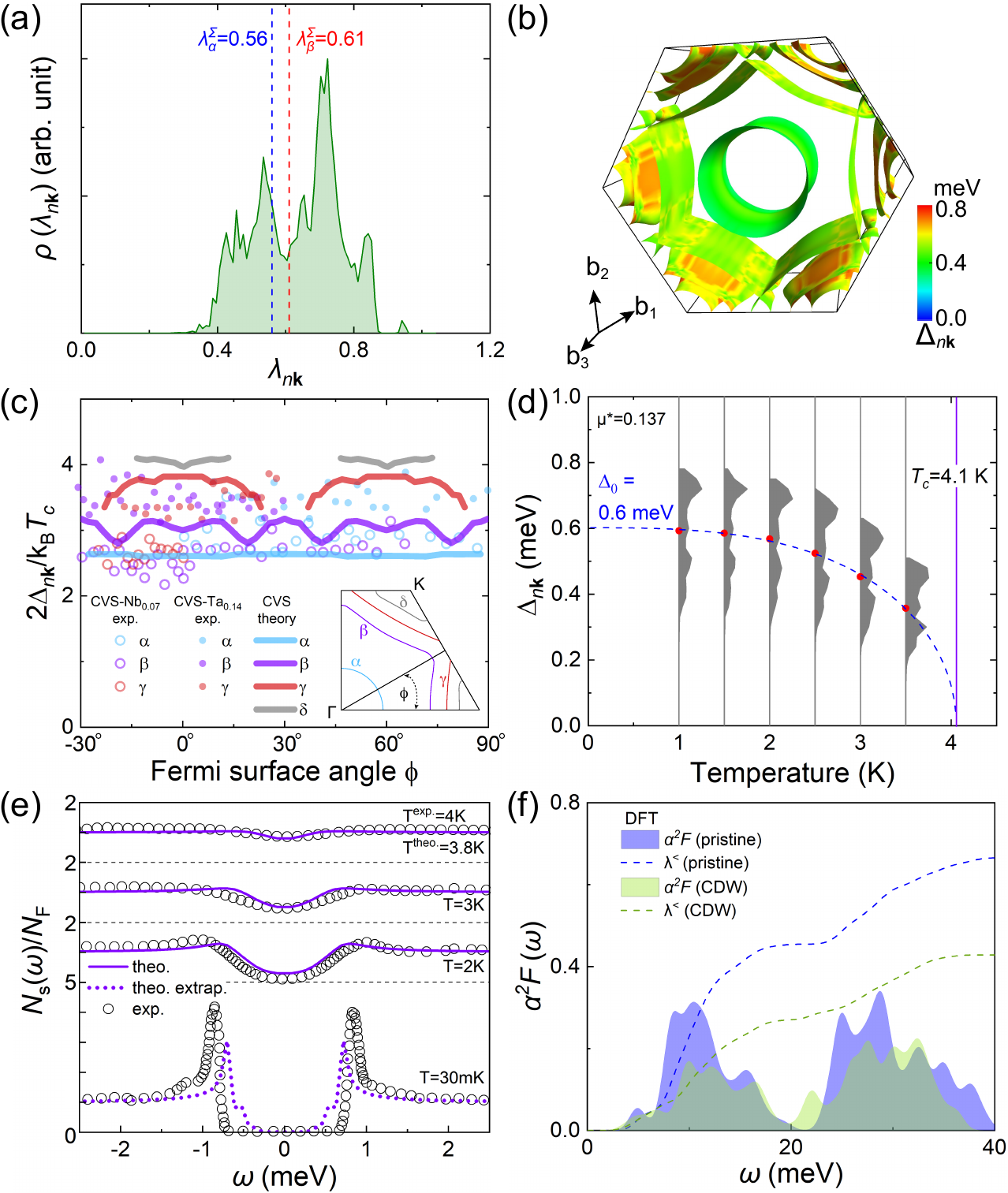}\\
  \caption{(a) Distribution density of $\lambda_{n {\rm \mathbf{k}}}$ near $E_\text{F}$, along with the kink-estimated $e$-ph coupling strength $\lambda^\Sigma$ from the $\alpha$- and $\beta$-bands. (b) Superconducting gaps $\Delta_{n\mathbf{k}}$ on the FSs computed by solving the fully anisotropic $GW$-Eliashberg equations with $\mu^{\ast}$ = 0.137 at $T$ = 1 K. (c) Angle distribution of the rescaled superconducting gaps $2\Delta_{n\textbf{k}}/k_B T_c$ on multiple FSs within the $k_z = 0$ plane, from our theoretical calculations of CsV$_3$Sb$_5$ (lines) and experimental data~\cite{Zhong2023a} of doped Cs(V$_{0.93}$Nb$_{0.07}$)$_3$Sb$_5$ (open circles) and Cs(V$_{0.86}$Ta$_{0.14}$)$_3$Sb$_5$ (solid dots). The angle $\phi$ is defined with respect to $k_x$ axis ($\Gamma-K$ direction) as depicted in the inset. (d) Distribution of superconducting gaps at various temperatures. The blue dashed line represents the temperature dependence of the averaged gap obtained by fitting to $\Delta(T)=\Delta_0[1-(T/T_c)^p]^{0.5}$, where $\Delta_0$ is the gap value at $T = 0$.
  (e) Calculated superconducting quasiparticle DOS $N_\text{s}/N_\text{F}$ as a function of temperature, compared with the experimental STS data~\cite{Deng2024}. The dotted line is a theoretical extrapolation to the low temperature (30 mK). (f) DFPT Eliashberg spectral functions and cumulative $e$-ph coupling for the pristine and CDW phases.
  }\label{fig4}
\end{figure}

Figure~\ref{fig4}(a) presents the density distribution of the electron-state resolved $e$-ph coupling $\lambda_{n\textbf{k}}$, showcasing a broad range from 0.3 to 0.9, indicating a complex distribution of $e$-ph coupling across different bands and the BZ. The kink-extracted $\lambda^\Sigma$ only represents certain parts of the whole $e$-ph landscape of CsV$_3$Sb$_5$. 
In the next, we proceed to compute superconducting properties by solving the anisotropic Eliashberg equations~\cite{Eliashberg1960,Allen1983,Margine2013} at the $GW$ level \cite{Li2024a}. 
In this work, we use an effective Coulomb repulsion pseudopotential $\mu^*$ = 0.137, which is estimated based on calculations of \textit{ab initio} DFT for superconductors (SCDFT)~\cite{Oliveira1988,Kawamura2020} (see Supplemental Material). 
Additionally, detailed analysis of the effect of different values of $\mu^*$ on the superconducting properties demonstrates that $T_c$ decreases progressively with increasing $\mu^*$ (see Fig. S7), as is expected. Within a physical range of $\mu^*$ values (0.175 $-$ 0.1), the superconducting $T_c$ (3.2 $-$ 5.2 K) remains consistent with experiments. These results (see Supplemental Material) demonstrate the appropriateness of adopting the value of $\mu^*$ = 0.137 to obtain the superconducting properties by solving Eliashberg equations.

Figure~\ref{fig4}(b) plots the distribution of the superconducting gap $\Delta_{n\mathbf{k}}$ (at $T$ = 1 K) on the Fermi surfaces (FSs), highlighting the nodeless and full-gap nature of the superconductivity. 
We further extract the superconducting gap magnitudes on the FSs within the $k_z=0$ plane, as shown in Fig.~\ref{fig4}(c). 
The angular distribution is relatively isotropic, with some variations around $M$ point ($\phi \sim 30^\circ$), complicated by the band crossings [see Figs.~\ref{fig1}(b) and (c)]. 
We compare our calculated angular distribution of $\Delta_{n\textbf{k}}$ with the ARPES-extracted gaps in doped CsV$_3$Sb$_5$ [Cs(V$_{0.93}$Nb$_{0.07}$)$_3$Sb$_5$ and Cs(V$_{0.86}$Ta$_{0.14}$)$_3$Sb$_5$] from Ref.~\cite{Zhong2023a}. 
Due to the differences in $T_c$ as well as the superconducting gap size for samples with different doping levels, here we directly compare the rescaled gap size $2\Delta_{n\textbf{k}}/k_B T_c$ in Fig.~\ref{fig4}(c). 
Across the three FS sheets of $\alpha$, $\beta$ and $\gamma$, our calculated ratio agrees consistently with the experimental values. 
Furthermore, we reveal strong superconducting gaps on the small FS pocket ($\delta$-sheet) near the $K$ point. 
Experimentally, the $\delta$-sheet was not observed~\cite{Zhong2023a}. One possibility is that in the CDW phase, the $\gamma$- and $\delta$-bands merge together and become indistinguishable near $E_\text{F}$, due to the band mixing and gap opening induced by the structural distortions [see Fig.~\ref{fig1}(d)].

Figure~\ref{fig4}(d) shows the distribution of the superconducting gaps at different temperatures. Our Eliashberg-equation calculations (with $\mu^*$ = 0.137) yield a $T_c$ = 4.1 K  and $\Delta_0$=0.6 meV (extracted value at $T$ = 0 K) for CsV$_3$Sb$_5$, agreeing well with the experimental $T_c \sim 2.5 - 5$ K~\cite{Chen2021,Liang2021,Xu2021,Duan2021,Gupta2022,Gupta2022a,He2022,Roppongi2023,Zhang2023} and gap size of $0.3 - 1.08$ meV \cite{Xu2021,Liang2021,Chen2021,Zhang2023,Gupta2022,Gupta2022a}.
Figure~\ref{fig4}(e) shows the computed superconducting quasiparticle DOS, $N_s(\omega)/N_{\rm F}$ (normalized by $N_\text{F}$, i.e., the DOS at $E_\text{F}$) at various temperatures, comparing with the experimental STS  spectra of Cs(V$_{0.86}$Ta$_{0.14}$)$_3$Sb$_5$~\cite{Deng2024}.
The agreement between the theoretical results and the experiments strongly supports the $e$-ph coupling as the origin for the superconductivity.

We now revisit the impact of CDW after comprehensively discussing the diverse manifestations of the $e$-ph coupling in the photoemission kinks and superconductivity of CsV$_3$Sb$_5$. 
Our DFPT calculations show that $\lambda$ decreases from 0.67 in the pristine phase to 0.43 in the CDW phase, as shown in Fig.~\ref{fig4}(f). 
The $e$-ph coupling constant $\lambda$ remains moderately strong and is consistent with the experimental estimation \cite{Zhong2023} of $\sim0.4$ (see Supplemental Material for more discussions). 
Clearly, CDW competes with the phonon-mediated superconductivity in CsV$_3$Sb$_5$. Notably, the strong $e$-ph coupled $\alpha$- and $\beta$- bands remain nearly intact within the CDW phase, suggesting the photoemission kinks in $\alpha$- and $\beta$-bands are less affected, in agreement with the experimental observations \cite{Zhong2023,Zhong2023a}. 
On a different note, spin fluctuations \cite{Kawamura2020} are found to weakly compete with the $e$-ph coupling in mediating the superconductivity in the pristine phase of CsV$_3$Sb$_5$ (see Supplemental Material), and do not alter the main conclusions of this work on the decisive role of $e$-ph coupling in superconductivity.

In summary, our systematic \textit{ab initio} study has established the unambiguous evidence that the $e$-ph coupling in CsV$_3$Sb$_5$ is responsible for both the multimodal photoemission kinks and the nodeless superconductivity. 
Our work establishes a faithful \textit{ab initio} many-body computational framework to resolve intertwined orders in complex materials, and especially clarifies the important and versatile role of $e$-ph coupling in shaping various low-energy electron excitations.

\textit{Acknowledgement.} This work was mainly supported by the Seed Fund provided by the Ershaghi Center for Energy Transition (E-CET) at  the Viterbi School of Engineering, University of Southern California (USC) (Z.L.). The Zumberge Preliminary Studies Research Award at USC provided support to J.Y.Y. and partially to Z.L. in this project.
M.D.B was supported by the Center for Computational Study of
Excited-State Phenomena in Energy Materials (C2SEPEM) at Lawrence Berkeley National Laboratory (LBNL), which is funded by the U.S.
Department of Energy (DOE), Office of Science, Basic Energy Sciences, Materials Sciences and Engineering
Division under Contract No. DEAC02-05CH11231, as part of the Computational Materials Sciences
Program. 
C.E.H. acknowledges the National Science and Technology Council “Ph.D. Students Study Abroad Program” of Taiwan.
Advanced codes were provided by C2SEPEM at LBNL. 
An award for computer time was provided by the U.S. DOE's  Innovative and Novel Computational Impact on Theory and Experiment (INCITE) Program. 
Computational resources for the $GW$ and $GW$PT calculations were provided by Frontier at the Oak Ridge Leadership Computing Facility, the Oak Ridge National Laboratory, which is supported by the Office of Science of the U.S. DOE under Contract No. DE-AC05-00OR22725.
Computational resources for the DFT, DFPT, and EPW calculations were provided by Frontera at Texas Advanced Computing Center, which is supported by National Science Foundation under Grant No. OAC-1818253.
Z.L. thanks Jiawei Ruan for helpful discussions.

\bibliography{ref}

\end{document}